\documentclass{article} % For LaTeX2e
\usepackage{nips14submit_e,times}
\usepackage{url}
\usepackage{amsmath}
\usepackage{amssymb}
\usepackage{graphicx}
\usepackage{natbib}
\usepackage{texments}

\title{Testing MCMC code}

\author{
Roger B. Grosse \\
Deptartment of Computer Science\\
University of Toronto \\
\texttt{rgrosse@cs.toronto.edu}
\And
David K. Duvenaud \\
School of Engineering and Applied Sciences \\
Harvard University \\
\texttt{dduvenaud@seas.harvard.edu}
}

\newcommand{\given}{\,|\,}

\nipsfinalcopy % Uncomment for camera-ready version

\begin{document}

\usestyle{default} %for texments

\maketitle

\begin{abstract}
Markov Chain Monte Carlo (MCMC) algorithms are a workhorse of probabilistic modeling and inference, but are difficult to debug, and are prone to silent failure if implemented na\"{\i}vely.
We outline several strategies for testing the correctness of MCMC algorithms.
Specifically, we advocate writing code in a modular way, where conditional probability calculations are kept separate from the logic of the sampler.
We discuss strategies for both unit testing and integration testing. As a running example, we show how a Python implementation of Gibbs sampling for a mixture of Gaussians model can be tested.\footnote{This tutorial is an adaptation of blog posts which can be found at \url{https://hips.seas.harvard.edu/blog/2013/05/20/testing-mcmc-code-part-1-unit-tests/} and \url{https://hips.seas.harvard.edu/blog/2013/06/10/testing-mcmc-code-part-2-integration-tests/}.}
\end{abstract}

\section{Introduction}

%When you write a nontrivial piece of software, how often do you get it completely correct on the first try?  When you implement a machine learning algorithm, how thorough are your tests?  If your answers are "rarely" and "not very," stop and think about the implications.

In machine learning, we often justify the correctness of our algorithms by showing that their mathematical idealizations satisfy certain properties. However, the algorithms must ultimately be implemented in software, so the question arises: how can we check that our code correctly implements the mathematical specification of an algorithm?  In this tutorial, we outline a set of techniques we have found useful for testing the correctness of one particular type of algorithm, a Markov chain Monte Carlo (MCMC) sampler. The techniques we outline are not novel, but we believe they deserve to be more widely known and practiced.

%There's a large literature on testing the convergence of optimization algorithms and MCMC samplers, but in this paper we'll talk about a more basic problem: how to test if your code correctly implements the mathematical specification of an algorithm. 
%There are a lot of general strategies software developers use to test code, but machine learning algorithms, and especially sampling algorithms, present their own difficulties:

Several factors conspire to make testing of MCMC code difficult:
\begin{itemize}
\item The algorithms are stochastic, so there's no single ``correct'' output.
\item Algorithms may perform badly for reasons other than buggy implementations, such as poor modeling assumptions or slow mixing between modes.
\item Good performance is often a matter of judgment: in a prediction task, if the algorithm correctly classifies 83.5\% of the test examples, does that mean it's working?
\end{itemize}

Furthermore, many machine learning algorithms have a curious property: they are robust against bugs \cite[\emph{e.g.}][]{conv-net-best-practices}.
Since they're designed to deal with noisy data, they can often compensate for noise caused by math mistakes as well.
A buggy algorithm might still make sensible-looking predictions.
This is problematic, because it means bugs can be subtle and hard to detect.
The algorithm might work well in some situations, such as small toy datasets used for validation, yet completely fail in other situations -- high dimensions, large numbers of training examples, noisy observations, etc.
Even if unnoticed bugs hurt performance by only a few percent, that difference may be important.

For researchers in particular, there is a further reason to be concerned about bugs, even if their effect on performance is negligible.
The job of a scientist isn't simply to write an algorithm that makes good predictions, but to run experiments which yield insight into the inner workings of an algorithm.
%Your job isn't just to write an algorithm that makes good predictions---it's to run experiments to figure out how one thing changes as a result of varying another.
Even if an algorithm performs well, implementation bugs can affect how one quantity varies as a result of changing another, for instance if some hyperparameter settings compensate for bugs better than others.
%Even if the algorithm ``works,'' it may have funny behaviors as some parameterizations compensate for mistakes better than others. 
%Bugs may cause curious results, such as performance worsening as you add more training data. Even worse, you may get plausible but incorrect results. Even if these aren't the experiments you eventually publish, they'll still mess up your intuition about the problem.

In this paper, we focus on testing MCMC samplers, partly because they are especially good illustrations of the challenges involved in testing machine learning code. 
%These strategies are presumably more widely applicable, though.
The particular practices we highlight are specific to MCMC, but they also exemplify useful strategies for testing implementations of machine learning algorithms.

 In software development, it is useful to distinguish two different kinds of tests: unit tests and integration tests. Unit tests are short and simple tests which check the correctness of small pieces of code, such as individual functions. The idea is for the tests to be as local as possible -- ideally, a single bug should cause a single test to fail. Integration tests test the overall behavior of the system, without reference to the underlying implementation. They are more global than unit tests: they test whether different parts of the system interact correctly to produce the desired behavior. Both kinds of tests are relevant to MCMC samplers. In particular, we discuss unit tests which check that conditional distributions are consistent with the joint distribution. We then review the Geweke test \cite{geweke-test}, which we view as a form of integration testing. We present these techniques using a Gibbs sampler as a running example, and discuss how MCMC code can be written in a modular structure which enables testing.

% A development pattern called test-driven development is built around testability: the idea is to first write unit and integration tests as a sort of formal specification for the software, and then to write code to make the tests pass. A fully test-driven development process might look something like:

% \begin{enumerate}
% \item Write integration tests based on a high-level specification of the system.
% \item Write unit tests.
% \item Write code which makes the unit tests pass.
% \item Make sure the integration tests pass.
% \end{enumerate}

% In research, because the requirements are so amorphous, it is often difficult to come up with good specifications in advance. However, we've found the general idea of getting unit tests to pass, followed by integration tests, to be a useful one.

\section{Example: mixture of Gaussians}
\label{sec:example}

As a running example throughout this tutorial, we will implement and test a Gibbs sampler for an isotropic mixture of Gaussians model in Python (with NumPy):
\begin{align*}
\pi &\sim {\rm Dirichlet}(\alpha) \\
\sigma^2_\mu &\sim {\rm InverseGamma}(a_\mu, b_\mu) \\
\sigma^2_n &\sim {\rm InverseGamma}(a_n, b_n) \\
z_i \given \pi &\sim {\rm Multinomial}(\pi) \\
\mu_{kj} \given \sigma^2_\mu &\sim {\rm Normal}(0, \sigma^2_\mu) \\
x_{ij} \given z_i, \mu_{z_i, j}, \sigma^2_n &\sim {\rm Normal}(\mu_{z_i, j}, \sigma^2_n)
\end{align*}

This is a toy model which is essentially a Bayesian analogue of K-means; however, the strategies presented here have served us well with more complex models and inference algorithms.
Here are the classes which define the model and the state (\emph{i.e.}~the variables which are sampled):

\begin{scriptsize}
\begin{Verbatim}[commandchars=\\\{\}]
\PY{k}{class} \PY{n+nc}{Model}\PY{p}{:}
    \PY{k}{def} \PY{n+nf}{\PYZus{}\PYZus{}init\PYZus{}\PYZus{}}\PY{p}{(}\PY{n+nb+bp}{self}\PY{p}{,} \PY{n}{alpha}\PY{p}{,} \PY{n}{K}\PY{p}{,} \PY{n}{sigma\PYZus{}sq\PYZus{}mu\PYZus{}prior}\PY{p}{,} \PY{n}{sigma\PYZus{}sq\PYZus{}n\PYZus{}prior}\PY{p}{)}\PY{p}{:}
        \PY{n+nb+bp}{self}\PY{o}{.}\PY{n}{alpha} \PY{o}{=} \PY{n}{alpha}             \PY{c}{\PYZsh{} Parameter for Dirichlet prior over mixture probabilities}
        \PY{n+nb+bp}{self}\PY{o}{.}\PY{n}{K} \PY{o}{=} \PY{n}{K}                     \PY{c}{\PYZsh{} Number of components}
        \PY{n+nb+bp}{self}\PY{o}{.}\PY{n}{sigma\PYZus{}sq\PYZus{}mu\PYZus{}prior} \PY{o}{=} \PY{n}{sigma\PYZus{}sq\PYZus{}mu\PYZus{}prior}
        \PY{n+nb+bp}{self}\PY{o}{.}\PY{n}{sigma\PYZus{}sq\PYZus{}n\PYZus{}prior} \PY{o}{=} \PY{n}{sigma\PYZus{}sq\PYZus{}n\PYZus{}prior}

\PY{k}{class} \PY{n+nc}{State}\PY{p}{:}
    \PY{k}{def} \PY{n+nf}{\PYZus{}\PYZus{}init\PYZus{}\PYZus{}}\PY{p}{(}\PY{n+nb+bp}{self}\PY{p}{,} \PY{n}{z}\PY{p}{,} \PY{n}{mu}\PY{p}{,} \PY{n}{sigma\PYZus{}sq\PYZus{}mu}\PY{p}{,} \PY{n}{sigma\PYZus{}sq\PYZus{}n}\PY{p}{,} \PY{n}{pi}\PY{p}{)}\PY{p}{:}
        \PY{n+nb+bp}{self}\PY{o}{.}\PY{n}{z} \PY{o}{=} \PY{n}{z}                     \PY{c}{\PYZsh{} Assignments (represented as an array of integers)}
        \PY{n+nb+bp}{self}\PY{o}{.}\PY{n}{mu} \PY{o}{=} \PY{n}{mu}                   \PY{c}{\PYZsh{} Cluster centers}
        \PY{n+nb+bp}{self}\PY{o}{.}\PY{n}{sigma\PYZus{}sq\PYZus{}mu} \PY{o}{=} \PY{n}{sigma\PYZus{}sq\PYZus{}mu} \PY{c}{\PYZsh{} Between\PYZhy{}cluster variance}
        \PY{n+nb+bp}{self}\PY{o}{.}\PY{n}{sigma\PYZus{}sq\PYZus{}n} \PY{o}{=} \PY{n}{sigma\PYZus{}sq\PYZus{}n}   \PY{c}{\PYZsh{} Within\PYZhy{}cluster variance}
        \PY{n+nb+bp}{self}\PY{o}{.}\PY{n}{pi} \PY{o}{=} \PY{n}{pi}                   \PY{c}{\PYZsh{} Mixture probabilities}
\end{Verbatim}
\end{scriptsize}

As a form of modularity, we define several distribution classes which know how to generate samples and evaluate log probabilities. For instance,

\begin{scriptsize}
\begin{Verbatim}[commandchars=\\\{\}]
\PY{k}{class} \PY{n+nc}{GaussianDistribution}\PY{p}{:}
    \PY{k}{def} \PY{n+nf}{\PYZus{}\PYZus{}init\PYZus{}\PYZus{}}\PY{p}{(}\PY{n+nb+bp}{self}\PY{p}{,} \PY{n}{mu}\PY{p}{,} \PY{n}{sigma\PYZus{}sq}\PY{p}{)}\PY{p}{:}
        \PY{n+nb+bp}{self}\PY{o}{.}\PY{n}{mu} \PY{o}{=} \PY{n}{mu}
        \PY{n+nb+bp}{self}\PY{o}{.}\PY{n}{sigma\PYZus{}sq} \PY{o}{=} \PY{n}{sigma\PYZus{}sq}

    \PY{k}{def} \PY{n+nf}{log\PYZus{}p}\PY{p}{(}\PY{n+nb+bp}{self}\PY{p}{,} \PY{n}{x}\PY{p}{)}\PY{p}{:}
        \PY{k}{return} \PY{o}{\PYZhy{}}\PY{l+m+mf}{0.5} \PY{o}{*} \PY{n}{np}\PY{o}{.}\PY{n}{log}\PY{p}{(}\PY{l+m+mi}{2}\PY{o}{*}\PY{n}{np}\PY{o}{.}\PY{n}{pi}\PY{p}{)} \PY{o}{+} \PYZbs{}
               \PY{o}{\PYZhy{}}\PY{l+m+mf}{0.5} \PY{o}{*} \PY{n}{np}\PY{o}{.}\PY{n}{log}\PY{p}{(}\PY{n+nb+bp}{self}\PY{o}{.}\PY{n}{sigma\PYZus{}sq}\PY{p}{)} \PY{o}{+} \PYZbs{}
               \PY{o}{\PYZhy{}}\PY{l+m+mf}{0.5} \PY{o}{*} \PY{p}{(}\PY{n}{x} \PY{o}{\PYZhy{}} \PY{n+nb+bp}{self}\PY{o}{.}\PY{n}{mu}\PY{p}{)} \PY{o}{*}\PY{o}{*} \PY{l+m+mi}{2} \PY{o}{/} \PY{n+nb+bp}{self}\PY{o}{.}\PY{n}{sigma\PYZus{}sq}

    \PY{k}{def} \PY{n+nf}{sample}\PY{p}{(}\PY{n+nb+bp}{self}\PY{p}{)}\PY{p}{:}
        \PY{k}{return} \PY{n}{np}\PY{o}{.}\PY{n}{random}\PY{o}{.}\PY{n}{normal}\PY{p}{(}\PY{n+nb+bp}{self}\PY{o}{.}\PY{n}{mu}\PY{p}{,} \PY{n}{np}\PY{o}{.}\PY{n}{sqrt}\PY{p}{(}\PY{n+nb+bp}{self}\PY{o}{.}\PY{n}{sigma\PYZus{}sq}\PY{p}{)}\PY{p}{)}
\end{Verbatim}
\end{scriptsize}

The following functions compute each of the conditional probability distributions required by the Gibbs sampler:

\begin{scriptsize}
\begin{Verbatim}[commandchars=\\\{\}]
\PY{k}{class} \PY{n+nc}{Model}\PY{p}{:}
    \PY{o}{.}\PY{o}{.}\PY{o}{.}

    \PY{k}{def} \PY{n+nf}{cond\PYZus{}pi}\PY{p}{(}\PY{n+nb+bp}{self}\PY{p}{,} \PY{n}{state}\PY{p}{)}\PY{p}{:}
        \PY{n}{counts} \PY{o}{=} \PY{n}{np}\PY{o}{.}\PY{n}{bincount}\PY{p}{(}\PY{n}{state}\PY{o}{.}\PY{n}{z}\PY{p}{)}
        \PY{n}{counts}\PY{o}{.}\PY{n}{resize}\PY{p}{(}\PY{n+nb+bp}{self}\PY{o}{.}\PY{n}{K}\PY{p}{)}
        \PY{k}{return} \PY{n}{DirichletDistribution}\PY{p}{(}\PY{n+nb+bp}{self}\PY{o}{.}\PY{n}{alpha} \PY{o}{+} \PY{n}{counts}\PY{p}{)}

    \PY{k}{def} \PY{n+nf}{cond\PYZus{}z}\PY{p}{(}\PY{n+nb+bp}{self}\PY{p}{,} \PY{n}{state}\PY{p}{,} \PY{n}{X}\PY{p}{)}\PY{p}{:}
        \PY{n}{nax} \PY{o}{=} \PY{n}{np}\PY{o}{.}\PY{n}{newaxis}
        \PY{n}{prior} \PY{o}{=} \PY{n}{np}\PY{o}{.}\PY{n}{log}\PY{p}{(}\PY{n}{state}\PY{o}{.}\PY{n}{pi}\PY{p}{)}
        \PY{n}{evidence} \PY{o}{=} \PY{n}{GaussianDistribution}\PY{p}{(}\PY{n}{state}\PY{o}{.}\PY{n}{mu}\PY{p}{[}\PY{n}{nax}\PY{p}{,} \PY{p}{:}\PY{p}{,} \PY{p}{:}\PY{p}{]}\PY{p}{,} \PY{n}{state}\PY{o}{.}\PY{n}{sigma\PYZus{}sq\PYZus{}n}\PY{p}{)}\PY{o}{.}\PY{n}{log\PYZus{}p}\PY{p}{(}\PY{n}{X}\PY{p}{[}\PY{p}{:}\PY{p}{,} \PY{n}{nax}\PY{p}{,} \PY{p}{:}\PY{p}{]}\PY{p}{)}\PY{o}{.}\PY{n}{sum}\PY{p}{(}\PY{l+m+mi}{2}\PY{p}{)}
        \PY{k}{return} \PY{n}{MultinomialDistribution}\PY{o}{.}\PY{n}{from\PYZus{}log\PYZus{}odds}\PY{p}{(}\PY{n}{prior}\PY{p}{[}\PY{n}{nax}\PY{p}{,} \PY{p}{:}\PY{p}{]} \PY{o}{+} \PY{n}{evidence}\PY{p}{)}

    \PY{k}{def} \PY{n+nf}{cond\PYZus{}mu}\PY{p}{(}\PY{n+nb+bp}{self}\PY{p}{,} \PY{n}{state}\PY{p}{,} \PY{n}{X}\PY{p}{)}\PY{p}{:}
        \PY{n}{ndata}\PY{p}{,} \PY{n}{ndim} \PY{o}{=} \PY{n}{X}\PY{o}{.}\PY{n}{shape}
        \PY{n}{h} \PY{o}{=} \PY{n}{np}\PY{o}{.}\PY{n}{zeros}\PY{p}{(}\PY{p}{(}\PY{n+nb+bp}{self}\PY{o}{.}\PY{n}{K}\PY{p}{,} \PY{n}{ndim}\PY{p}{)}\PY{p}{)}
        \PY{n}{lam} \PY{o}{=} \PY{n}{np}\PY{o}{.}\PY{n}{zeros}\PY{p}{(}\PY{p}{(}\PY{n+nb+bp}{self}\PY{o}{.}\PY{n}{K}\PY{p}{,} \PY{n}{ndim}\PY{p}{)}\PY{p}{)}
        \PY{k}{for} \PY{n}{k} \PY{o+ow}{in} \PY{n+nb}{range}\PY{p}{(}\PY{n+nb+bp}{self}\PY{o}{.}\PY{n}{K}\PY{p}{)}\PY{p}{:}
            \PY{n}{idxs} \PY{o}{=} \PY{n}{np}\PY{o}{.}\PY{n}{where}\PY{p}{(}\PY{n}{state}\PY{o}{.}\PY{n}{z} \PY{o}{==} \PY{n}{k}\PY{p}{)}\PY{p}{[}\PY{l+m+mi}{0}\PY{p}{]}
            \PY{k}{if} \PY{n}{idxs}\PY{o}{.}\PY{n}{size} \PY{o}{\PYZgt{}} \PY{l+m+mi}{0}\PY{p}{:}
                \PY{n}{h}\PY{p}{[}\PY{n}{k}\PY{p}{,} \PY{p}{:}\PY{p}{]} \PY{o}{=} \PY{n}{X}\PY{p}{[}\PY{n}{idxs}\PY{p}{,} \PY{p}{:}\PY{p}{]}\PY{o}{.}\PY{n}{sum}\PY{p}{(}\PY{l+m+mi}{0}\PY{p}{)} \PY{o}{/} \PY{n}{state}\PY{o}{.}\PY{n}{sigma\PYZus{}sq\PYZus{}n}
                \PY{n}{lam}\PY{p}{[}\PY{n}{k}\PY{p}{,} \PY{p}{:}\PY{p}{]} \PY{o}{=} \PY{n}{idxs}\PY{o}{.}\PY{n}{size} \PY{o}{/} \PY{n}{state}\PY{o}{.}\PY{n}{sigma\PYZus{}sq\PYZus{}n} \PY{o}{+} \PY{l+m+mf}{1.} \PY{o}{/} \PY{n}{state}\PY{o}{.}\PY{n}{sigma\PYZus{}sq\PYZus{}mu}
            \PY{k}{else}\PY{p}{:}
                \PY{n}{h}\PY{p}{[}\PY{n}{k}\PY{p}{,} \PY{p}{:}\PY{p}{]} \PY{o}{=} \PY{l+m+mf}{0.}
                \PY{n}{lam}\PY{p}{[}\PY{n}{k}\PY{p}{,} \PY{p}{:}\PY{p}{]} \PY{o}{=} \PY{l+m+mf}{1.} \PY{o}{/} \PY{n}{state}\PY{o}{.}\PY{n}{sigma\PYZus{}sq\PYZus{}mu}
        \PY{k}{return} \PY{n}{GaussianDistribution}\PY{p}{(}\PY{n}{h} \PY{o}{/} \PY{n}{lam}\PY{p}{,} \PY{l+m+mf}{1.} \PY{o}{/} \PY{n}{lam}\PY{p}{)}

    \PY{k}{def} \PY{n+nf}{cond\PYZus{}sigma\PYZus{}sq\PYZus{}mu}\PY{p}{(}\PY{n+nb+bp}{self}\PY{p}{,} \PY{n}{state}\PY{p}{)}\PY{p}{:}
        \PY{n}{ndim} \PY{o}{=} \PY{n}{state}\PY{o}{.}\PY{n}{mu}\PY{o}{.}\PY{n}{shape}\PY{p}{[}\PY{l+m+mi}{1}\PY{p}{]}
        \PY{n}{a} \PY{o}{=} \PY{n+nb+bp}{self}\PY{o}{.}\PY{n}{sigma\PYZus{}sq\PYZus{}mu\PYZus{}prior}\PY{o}{.}\PY{n}{a} \PY{o}{+} \PYZbs{}
            \PY{l+m+mf}{0.5} \PY{o}{*} \PY{n+nb+bp}{self}\PY{o}{.}\PY{n}{K} \PY{o}{*} \PY{n}{ndim}
        \PY{n}{b} \PY{o}{=} \PY{n+nb+bp}{self}\PY{o}{.}\PY{n}{sigma\PYZus{}sq\PYZus{}mu\PYZus{}prior}\PY{o}{.}\PY{n}{b} \PY{o}{+} \PYZbs{}
            \PY{l+m+mf}{0.5} \PY{o}{*} \PY{n}{np}\PY{o}{.}\PY{n}{sum}\PY{p}{(}\PY{n}{state}\PY{o}{.}\PY{n}{mu} \PY{o}{*}\PY{o}{*} \PY{l+m+mi}{2}\PY{p}{)}
        \PY{k}{return} \PY{n}{InverseGammaDistribution}\PY{p}{(}\PY{n}{a}\PY{p}{,} \PY{n}{b}\PY{p}{)}

    \PY{k}{def} \PY{n+nf}{cond\PYZus{}sigma\PYZus{}sq\PYZus{}n}\PY{p}{(}\PY{n+nb+bp}{self}\PY{p}{,} \PY{n}{state}\PY{p}{,} \PY{n}{X}\PY{p}{)}\PY{p}{:}
        \PY{n}{ndata}\PY{p}{,} \PY{n}{ndim} \PY{o}{=} \PY{n}{X}\PY{o}{.}\PY{n}{shape}
        \PY{n}{a} \PY{o}{=} \PY{n+nb+bp}{self}\PY{o}{.}\PY{n}{sigma\PYZus{}sq\PYZus{}n\PYZus{}prior}\PY{o}{.}\PY{n}{a} \PY{o}{+} \PYZbs{}
            \PY{l+m+mf}{0.5} \PY{o}{*} \PY{n}{ndata} \PY{o}{*} \PY{n}{ndim}
        \PY{n}{b} \PY{o}{=} \PY{n+nb+bp}{self}\PY{o}{.}\PY{n}{sigma\PYZus{}sq\PYZus{}n\PYZus{}prior}\PY{o}{.}\PY{n}{b} \PY{o}{+} \PYZbs{}
            \PY{l+m+mf}{0.5} \PY{o}{*} \PY{n}{np}\PY{o}{.}\PY{n}{sum}\PY{p}{(}\PY{p}{(}\PY{n}{X} \PY{o}{\PYZhy{}} \PY{n}{state}\PY{o}{.}\PY{n}{mu}\PY{p}{[}\PY{n}{state}\PY{o}{.}\PY{n}{z}\PY{p}{,} \PY{p}{:}\PY{p}{]}\PY{p}{)} \PY{o}{*}\PY{o}{*} \PY{l+m+mi}{2}\PY{p}{)}
        \PY{k}{return} \PY{n}{InverseGammaDistribution}\PY{p}{(}\PY{n}{a}\PY{p}{,} \PY{n}{b}\PY{p}{)}
\end{Verbatim}
\end{scriptsize}

Finally, the Gibbs sampling routine itself is a simple wrapper around the conditional probability distributions:

\begin{scriptsize}
\begin{Verbatim}[commandchars=\\\{\}]
\PY{k}{class} \PY{n+nc}{Model}\PY{p}{:}
    \PY{o}{.}\PY{o}{.}\PY{o}{.}

    \PY{k}{def} \PY{n+nf}{gibbs\PYZus{}step}\PY{p}{(}\PY{n+nb+bp}{self}\PY{p}{,} \PY{n}{state}\PY{p}{,} \PY{n}{X}\PY{p}{)}\PY{p}{:}
        \PY{n}{state}\PY{o}{.}\PY{n}{pi} \PY{o}{=} \PY{n+nb+bp}{self}\PY{o}{.}\PY{n}{cond\PYZus{}pi}\PY{p}{(}\PY{n}{state}\PY{p}{)}\PY{o}{.}\PY{n}{sample}\PY{p}{(}\PY{p}{)}
        \PY{n}{state}\PY{o}{.}\PY{n}{z} \PY{o}{=} \PY{n+nb+bp}{self}\PY{o}{.}\PY{n}{cond\PYZus{}z}\PY{p}{(}\PY{n}{state}\PY{p}{,} \PY{n}{X}\PY{p}{)}\PY{o}{.}\PY{n}{sample}\PY{p}{(}\PY{p}{)}
        \PY{n}{state}\PY{o}{.}\PY{n}{mu} \PY{o}{=} \PY{n+nb+bp}{self}\PY{o}{.}\PY{n}{cond\PYZus{}mu}\PY{p}{(}\PY{n}{state}\PY{p}{,} \PY{n}{X}\PY{p}{)}\PY{o}{.}\PY{n}{sample}\PY{p}{(}\PY{p}{)}
        \PY{n}{state}\PY{o}{.}\PY{n}{sigma\PYZus{}sq\PYZus{}mu} \PY{o}{=} \PY{n+nb+bp}{self}\PY{o}{.}\PY{n}{cond\PYZus{}sigma\PYZus{}sq\PYZus{}mu}\PY{p}{(}\PY{n}{state}\PY{p}{)}\PY{o}{.}\PY{n}{sample}\PY{p}{(}\PY{p}{)}
        \PY{n}{state}\PY{o}{.}\PY{n}{sigma\PYZus{}sq\PYZus{}n} \PY{o}{=} \PY{n+nb+bp}{self}\PY{o}{.}\PY{n}{cond\PYZus{}sigma\PYZus{}sq\PYZus{}n}\PY{p}{(}\PY{n}{state}\PY{p}{,} \PY{n}{X}\PY{p}{)}\PY{o}{.}\PY{n}{sample}\PY{p}{(}\PY{p}{)}
\end{Verbatim}
\end{scriptsize}

In the next section, we will discuss our motivations for decomposing the functionality in this particular way.

\section{Unit testing}

Unit testing is about independently testing small chunks of code. 
%Naturally, if you want to unit test your code, it needs to be divided into independently testable chunks, i.e.\ it needs to be modular. 
Therefore, the code must be written in a modular way, such that different chunks are independently testable.
Such a modular design doesn't happen automatically, and it can be difficult to retrofit a code base with unit tests. 
%In fact, the book \emph{Working effectively with legacy code} \cite{feathers2004working} defines legacy code as ``code without unit tests.'' The whole book is about strategies for refactoring an existing code base to make it testable.
(In fact, strategies for retrofitting are the subject of a whole book, \emph{Working effectively with legacy code} \cite{feathers2004working}, which defines legacy code as ``code without unit tests.'')
It's much easier to try to keep things modular and testable from the beginning. 
%In fact, one of the arguments behind test-driven development is that, by writing unit tests along with the actual code, you force yourself to structure your code in a modular way, and this winds up having other benefits like reusability and maintainability. This agrees with our experience.

Unfortunately, the way machine learning is typically taught encourages a programming style which is neither modular nor testable. In problem sets, students typically first derive the update rules for an iterative algorithm, and then implement the updates directly. This approach makes it easy for graders to spot differences from the correct solution, but it's not a robust way to build real software projects. Iterative algorithms can be difficult to test as black boxes.

Fortunately, most machine learning algorithms are formulated in terms of general mathematical principles that suggest a modular organization into testable chunks. For instance, we often formulate algorithms as optimization problems which can be solved using general purpose algorithms such as gradient descent or L-BFGS \cite{liu1989limited}. From an implementation standpoint, this requires writing functions to (a) evaluate the objective function at a point, and (b) compute the gradient of the objective function with respect to the model parameters. These functions are then fed to library optimization routines. The gradients can be checked using finite difference techniques. Conveniently, most scientific computing environments provide implementations of these checks; for instance, \verb+scipy.optimize.check_grad+ (in Python).

This organization into gradient computations and general purpose optimization routines exemplifies a useful design principle for machine learning code: separate out the implementations of the model and general-purpose algorithms. Consider how this principle can be applied in the context of MCMC. Seemingly the most straightforward way to implement a Gibbs sampler would be to derive by hand the update rules for the individual variables and then write a single iterative routine using those rules. As mentioned above, such an approach can be difficult to test.However, a Gibbs sampler is defined in terms of conditional probability calculations, and these calculations can be decomposed out.

In each update, we sample a variable $x$ from its conditional distribution $p(x|z)$. It is difficult to directly test the correctness of samples from the conditional distribution: such a test would likely be slow and nondeterministic, both highly undesirable properties for a unit test to have. We instead recommend testing that the conditional distribution is consistent with the joint distribution. In particular, for any two values $x$ and $x^\prime$, we must have that:
\begin{equation}
\frac{p(x^\prime | z)}{p(x | z)} = \frac{p(x^\prime, z)}{p(x, z)} \label{eqn:consistency}
\end{equation}
Importantly, this relationship must hold \emph{exactly} for any two values $x$ and $x^\prime$. For most models we use, if our formula for the conditional probability is wrong, then this relationship would likely fail for two randomly chosen values. Therefore, a simple and highly effective test for conditional probability computations is to choose random values for $x$, $x^\prime$, and $z$, and verify the above equality to a suitable precision, such as $10^{-10}$. (As usual in probabilistic inference, we should use log probabilities rather than probabilities for numerical reasons.)

In terms of implementation, this form of testing suggests writing three separate modules (as we did in Section~\ref{sec:example}):

\begin{enumerate}
\item Classes representing probability distributions. These classes should know how to (a) evaluate the log probability density function (or probability mass function) at a point, and (b) generate samples.
\item A specification of the model in terms of functions which compute the probability of a joint assignment and functions which return conditional probability distributions.
\item A Gibbs sampling routine, which sweeps over the variables in the model, replacing each one with a sample from its conditional distribution.
\end{enumerate}

These distribution classes require their own forms of unit testing. For instance, we may generate large numbers of samples from the distributions, and then ensure that the empirical moments are consistent with the exact moments.

This modular organization supports not just testability, but also reusability: it's easy to reuse the distribution classes between entirely different projects, and changing the model might require modifying only a few conditional distribution routines. Sophisticated Metropolis-Hastings proposals are often defined in terms of conditional probabilities, so much of the implementation can be shared.

%While Gibbs sampling is a relatively simple example, this thought process is a good way of coming up with modular and testable ways to implement machine learning algorithms. With more work, it can be extended to trickier samplers such as Metropolis-Hastings. The general idea is to make a list of mathematical properties that the component functions should satisfy, and then check that they actually satisfy those properties.

There are a lot of helpful software tools for managing unit tests. We use \verb nose \footnote{Available at \url{http://nose.readthedocs.org}}, which is a simple and elegant testing framework for Python. In order to avoid clutter, we keep the tests in a separate directory whose structure parallels the main code directory.

\subsection{Mixture of Gaussians example}

In order to unit test the Gibbs sampler of Section \ref{sec:example}, we must first write one additional function which computes the joint log probability of all the variables:

\begin{scriptsize}
\begin{Verbatim}[commandchars=\\\{\}]
\PY{k}{class} \PY{n+nc}{Model}\PY{p}{:}
    \PY{o}{.}\PY{o}{.}\PY{o}{.}

    \PY{k}{def} \PY{n+nf}{joint\PYZus{}log\PYZus{}p}\PY{p}{(}\PY{n+nb+bp}{self}\PY{p}{,} \PY{n}{state}\PY{p}{,} \PY{n}{X}\PY{p}{)}\PY{p}{:}
        \PY{k}{return} \PY{n}{DirichletDistribution}\PY{p}{(}\PY{n+nb+bp}{self}\PY{o}{.}\PY{n}{alpha} \PY{o}{*} \PY{n}{np}\PY{o}{.}\PY{n}{ones}\PY{p}{(}\PY{n+nb+bp}{self}\PY{o}{.}\PY{n}{K}\PY{p}{)}\PY{p}{)}\PY{o}{.}\PY{n}{log\PYZus{}p}\PY{p}{(}\PY{n}{state}\PY{o}{.}\PY{n}{pi}\PY{p}{)} \PY{o}{+} \PYZbs{}
               \PY{n}{MultinomialDistribution}\PY{o}{.}\PY{n}{from\PYZus{}probabilities}\PY{p}{(}\PY{n}{state}\PY{o}{.}\PY{n}{pi}\PY{p}{)}\PY{o}{.}\PY{n}{log\PYZus{}p}\PY{p}{(}\PY{n}{state}\PY{o}{.}\PY{n}{z}\PY{p}{)}\PY{o}{.}\PY{n}{sum}\PY{p}{(}\PY{p}{)} \PY{o}{+} \PYZbs{}
               \PY{n+nb+bp}{self}\PY{o}{.}\PY{n}{sigma\PYZus{}sq\PYZus{}mu\PYZus{}prior}\PY{o}{.}\PY{n}{log\PYZus{}p}\PY{p}{(}\PY{n}{state}\PY{o}{.}\PY{n}{sigma\PYZus{}sq\PYZus{}mu}\PY{p}{)} \PY{o}{+} \PYZbs{}
               \PY{n+nb+bp}{self}\PY{o}{.}\PY{n}{sigma\PYZus{}sq\PYZus{}n\PYZus{}prior}\PY{o}{.}\PY{n}{log\PYZus{}p}\PY{p}{(}\PY{n}{state}\PY{o}{.}\PY{n}{sigma\PYZus{}sq\PYZus{}n}\PY{p}{)} \PY{o}{+} \PYZbs{}
               \PY{n}{GaussianDistribution}\PY{p}{(}\PY{l+m+mf}{0.}\PY{p}{,} \PY{n}{state}\PY{o}{.}\PY{n}{sigma\PYZus{}sq\PYZus{}mu}\PY{p}{)}\PY{o}{.}\PY{n}{log\PYZus{}p}\PY{p}{(}\PY{n}{state}\PY{o}{.}\PY{n}{mu}\PY{p}{)}\PY{o}{.}\PY{n}{sum}\PY{p}{(}\PY{p}{)} \PY{o}{+} \PYZbs{}
               \PY{n}{GaussianDistribution}\PY{p}{(}\PY{n}{state}\PY{o}{.}\PY{n}{mu}\PY{p}{[}\PY{n}{state}\PY{o}{.}\PY{n}{z}\PY{p}{,} \PY{p}{:}\PY{p}{]}\PY{p}{,} \PY{n}{state}\PY{o}{.}\PY{n}{sigma\PYZus{}sq\PYZus{}n}\PY{p}{)}\PY{o}{.}\PY{n}{log\PYZus{}p}\PY{p}{(}\PY{n}{X}\PY{p}{)}\PY{o}{.}\PY{n}{sum}\PY{p}{(}\PY{p}{)}
\end{Verbatim}
\end{scriptsize}

We then test each of these conditional distributions in turn by checking the identity (\ref{eqn:consistency}). For instance,

\begin{scriptsize}
\begin{Verbatim}[commandchars=\\\{\}]
\PY{k}{def} \PY{n+nf}{test\PYZus{}cond\PYZus{}mu}\PY{p}{(}\PY{p}{)}\PY{p}{:}
    \PY{n}{model} \PY{o}{=} \PY{n}{random\PYZus{}model}\PY{p}{(}\PY{p}{)}
    \PY{n}{state}\PY{p}{,} \PY{n}{X} \PY{o}{=} \PY{n}{model}\PY{o}{.}\PY{n}{forward\PYZus{}sample}\PY{p}{(}\PY{n}{N}\PY{p}{,} \PY{n}{D}\PY{p}{)}
    \PY{n}{new\PYZus{}state} \PY{o}{=} \PY{n}{copy}\PY{o}{.}\PY{n}{deepcopy}\PY{p}{(}\PY{n}{state}\PY{p}{)}
    \PY{n}{new\PYZus{}state}\PY{o}{.}\PY{n}{mu} \PY{o}{=} \PY{n}{np}\PY{o}{.}\PY{n}{random}\PY{o}{.}\PY{n}{normal}\PY{p}{(}\PY{n}{size}\PY{o}{=}\PY{p}{(}\PY{n}{K}\PY{p}{,} \PY{n}{D}\PY{p}{)}\PY{p}{)}
    \PY{n}{cond} \PY{o}{=} \PY{n}{model}\PY{o}{.}\PY{n}{cond\PYZus{}mu}\PY{p}{(}\PY{n}{state}\PY{p}{,} \PY{n}{X}\PY{p}{)}
    \PY{k}{assert} \PY{n}{np}\PY{o}{.}\PY{n}{allclose}\PY{p}{(}\PY{n}{cond}\PY{o}{.}\PY{n}{log\PYZus{}p}\PY{p}{(}\PY{n}{new\PYZus{}state}\PY{o}{.}\PY{n}{mu}\PY{p}{)}\PY{o}{.}\PY{n}{sum}\PY{p}{(}\PY{p}{)} \PY{o}{\PYZhy{}} \PY{n}{cond}\PY{o}{.}\PY{n}{log\PYZus{}p}\PY{p}{(}\PY{n}{state}\PY{o}{.}\PY{n}{mu}\PY{p}{)}\PY{o}{.}\PY{n}{sum}\PY{p}{(}\PY{p}{)}\PY{p}{,}
                       \PY{n}{model}\PY{o}{.}\PY{n}{joint\PYZus{}log\PYZus{}p}\PY{p}{(}\PY{n}{new\PYZus{}state}\PY{p}{,} \PY{n}{X}\PY{p}{)} \PY{o}{\PYZhy{}} \PY{n}{model}\PY{o}{.}\PY{n}{joint\PYZus{}log\PYZus{}p}\PY{p}{(}\PY{n}{state}\PY{p}{,} \PY{n}{X}\PY{p}{)}\PY{p}{)}
\end{Verbatim}
\end{scriptsize}

It is a good idea to sanity check our tests by introducing a small bug and verifying that the test fails. Let's substitute $0.51$ for $0.5$ in our within-cluster variance computation:

\begin{scriptsize}
\begin{Verbatim}[commandchars=\\\{\}]
\PY{k}{class} \PY{n+nc}{Model}\PY{p}{:}
    \PY{o}{.}\PY{o}{.}\PY{o}{.}

    \PY{k}{def} \PY{n+nf}{cond\PYZus{}sigma\PYZus{}sq\PYZus{}n}\PY{p}{(}\PY{n+nb+bp}{self}\PY{p}{,} \PY{n}{state}\PY{p}{,} \PY{n}{X}\PY{p}{)}\PY{p}{:}
        \PY{n}{ndata}\PY{p}{,} \PY{n}{ndim} \PY{o}{=} \PY{n}{X}\PY{o}{.}\PY{n}{shape}
        \PY{n}{a} \PY{o}{=} \PY{n+nb+bp}{self}\PY{o}{.}\PY{n}{sigma\PYZus{}sq\PYZus{}n\PYZus{}prior}\PY{o}{.}\PY{n}{a} \PY{o}{+} \PYZbs{}
            \PY{l+m+mf}{0.5} \PY{o}{*} \PY{n}{ndata} \PY{o}{*} \PY{n}{ndim}
        \PY{n}{b} \PY{o}{=} \PY{n+nb+bp}{self}\PY{o}{.}\PY{n}{sigma\PYZus{}sq\PYZus{}n\PYZus{}prior}\PY{o}{.}\PY{n}{b} \PY{o}{+} \PYZbs{}
            \PY{l+m+mf}{0.51} \PY{o}{*} \PY{n}{np}\PY{o}{.}\PY{n}{sum}\PY{p}{(}\PY{p}{(}\PY{n}{X} \PY{o}{\PYZhy{}} \PY{n}{state}\PY{o}{.}\PY{n}{mu}\PY{p}{[}\PY{n}{state}\PY{o}{.}\PY{n}{z}\PY{p}{,} \PY{p}{:}\PY{p}{]}\PY{p}{)} \PY{o}{*}\PY{o}{*} \PY{l+m+mi}{2}\PY{p}{)}   \PY{c}{\PYZsh{} oops!}
        \PY{k}{return} \PY{n}{InverseGammaDistribution}\PY{p}{(}\PY{n}{a}\PY{p}{,} \PY{n}{b}\PY{p}{)}
\end{Verbatim}
\end{scriptsize}

Sure enough, \verb+nose+ catches it. Furthermore, the only unit test which fails is the one corresponding to this function:

\begin{scriptsize}
\begin{verbatim}
rgrosse:~/code/testing$ nosetests
....F
======================================================================
FAIL: test_mog.test_cond_sigma_sq_n
----------------------------------------------------------------------
Traceback (most recent call last):
  File "/Users/rgrosse/anaconda/lib/python2.7/site-packages/nose/case.py", line 197, in runTest
    self.test(*self.arg)
  File "/Users/rgrosse/code/testing/test_mog.py", line 74, in test_cond_sigma_sq_n
    model.joint_log_p(new_state, X) - model.joint_log_p(state, X))
AssertionError

----------------------------------------------------------------------
Ran 5 tests in 0.217s

FAILED (failures=1)
\end{verbatim}
\end{scriptsize}

If we repair \verb+cond_sigma_sq_n+, our tests pass:

\begin{scriptsize}
\begin{verbatim}
rgrosse:~/code/testing$ nosetests
.....
----------------------------------------------------------------------
Ran 5 tests in 0.231s

OK
\end{verbatim}
\end{scriptsize}

\section{Integration testing: the Geweke test}

Most of the mistakes we've caught in our own code were caught by unit tests. (Naturally, we have no idea about the ones we haven't caught.) But no matter how thoroughly we unit test, there are still subtle bugs that slip through the cracks. Integration testing is a more global approach, and tests the overall behavior of the software, which depends on the interaction of multiple components.

When testing MCMC samplers, we are interested in testing two things: whether our algorithm is mathematically correct, and whether the Markov chain has mixed. These two goals are somewhat independent of each other: a mathematically correct algorithm can get stuck and fail to find a good solution, and (counterintuitively) a mathematically incorrect algorithm can often find a good solution or at least get close. This section is about checking mathematical correctness.

One powerful technique for testing MCMC algorithms is the \emph{Geweke test} \cite{geweke-test}. The basic idea is simple: suppose we have a generative model over parameters $\theta$ and data $x$, and we want to test an MCMC sampler for the posterior $p(\theta | x)$. There are two different ways to sample from the joint distribution $p(\theta, x)$. First, we can forward sample, i.e.\ sample $\theta$ from $p(\theta)$, and then sample $x$ from $p(x | \theta)$. Second, we can start from a forward sample, and then run a Markov chain which alternates between

\begin{enumerate}
\item updating $\theta$ using one of the MCMC transition operators, which should preserve the distribution $p(\theta | x) $, and
\item resampling the data from the distribution $p(x | \theta) $.
\end{enumerate}

Since each of these operations preserves the joint distribution $p(\theta, x)$, each step of this chain is a perfect sample from $p(\theta, x)$.

If the sampler is correct, then each of these two procedures should yield samples from exactly the same distribution. We can test this by comparing the distributions of a variety of statistics of the samples, \emph{e.g.}~the mean of $x$ or the maximum absolute value of $\theta$. No matter which statistics we choose, the two distributions should be indistinguishable.

Geweke \cite{geweke-test} uses frequentist hypothesis tests to compare the distributions. This can be difficult, however, since one needs to account for dependencies between the samples in order to determine significance. We adopt a less principled but much simpler approach of looking at P-P plots of the statistics. Here are a few ways the plot could turn out:
\renewcommand{\tabcolsep}{0pt}
\begin{center}
\begin{tabular}{ccc}
Passes Geweke test & Fails Geweke test & Unclear \\
\includegraphics[width=0.32 \textwidth]{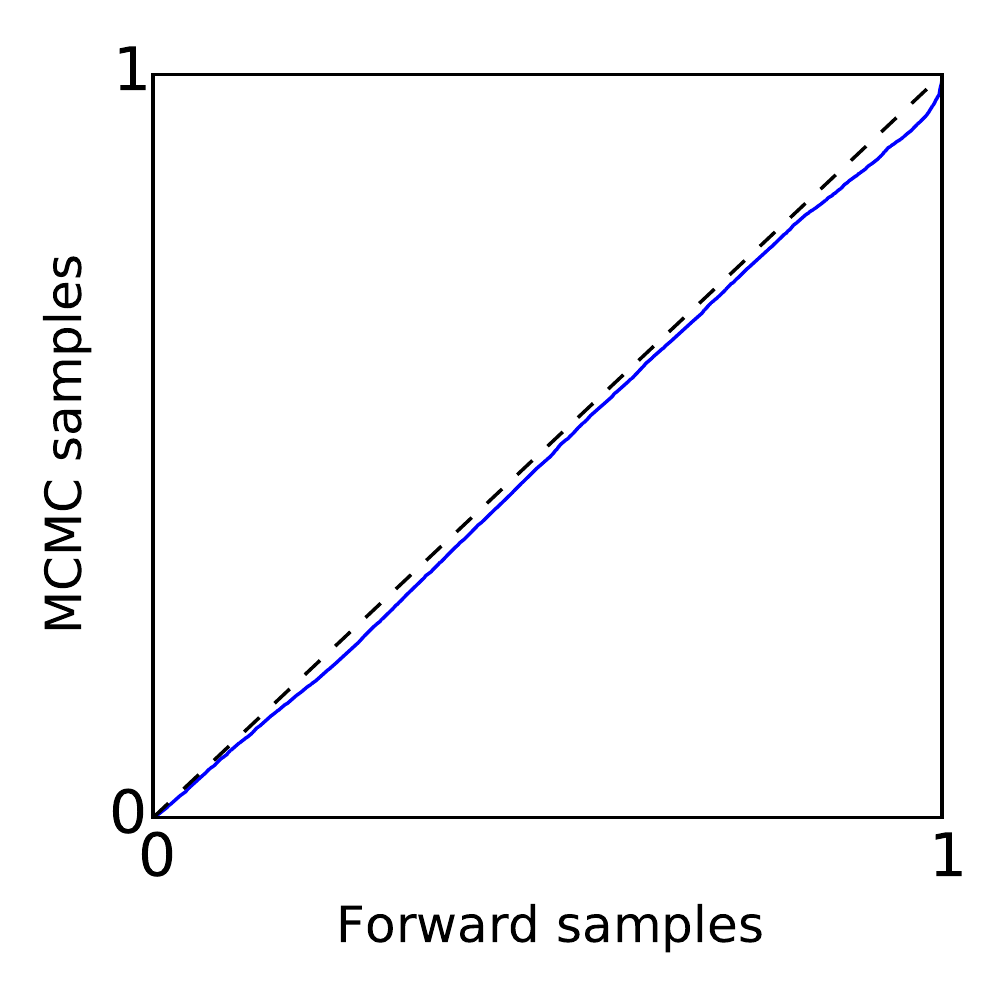} &
\includegraphics[width=0.32 \textwidth]{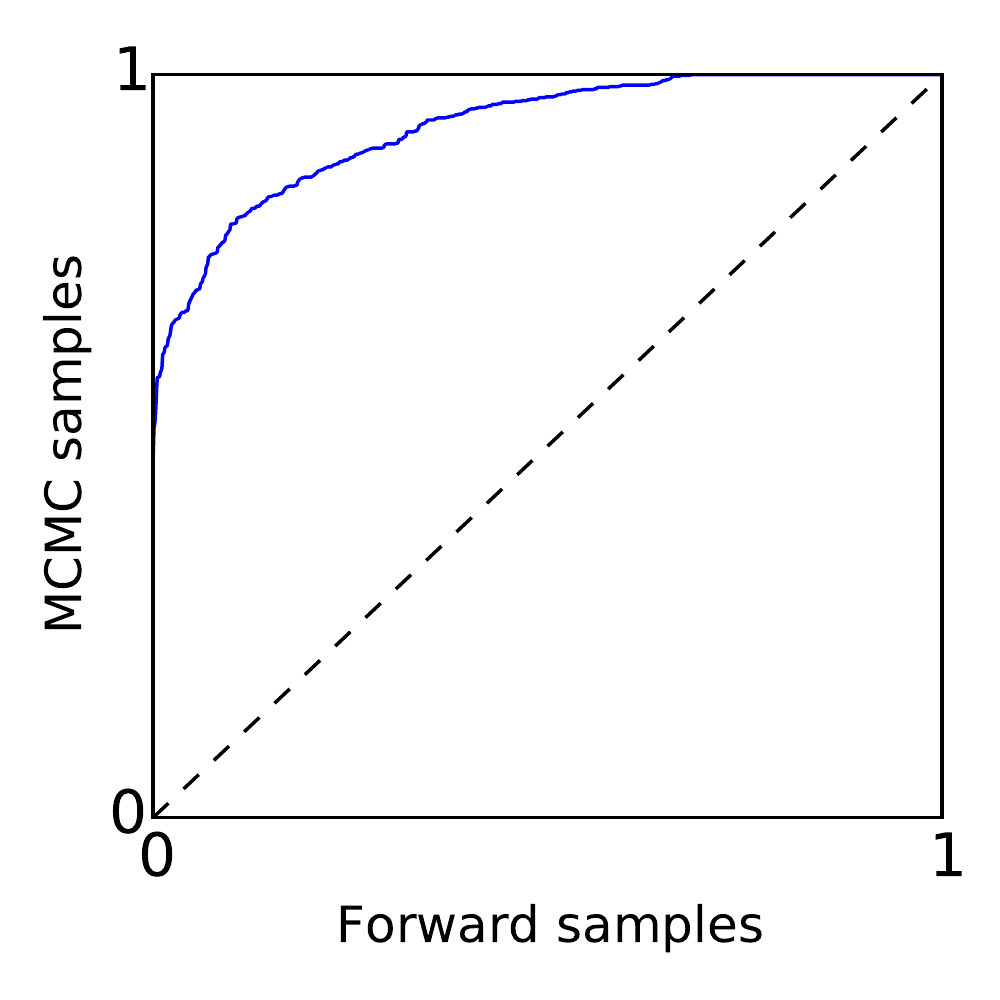} &
\includegraphics[width=0.32 \textwidth]{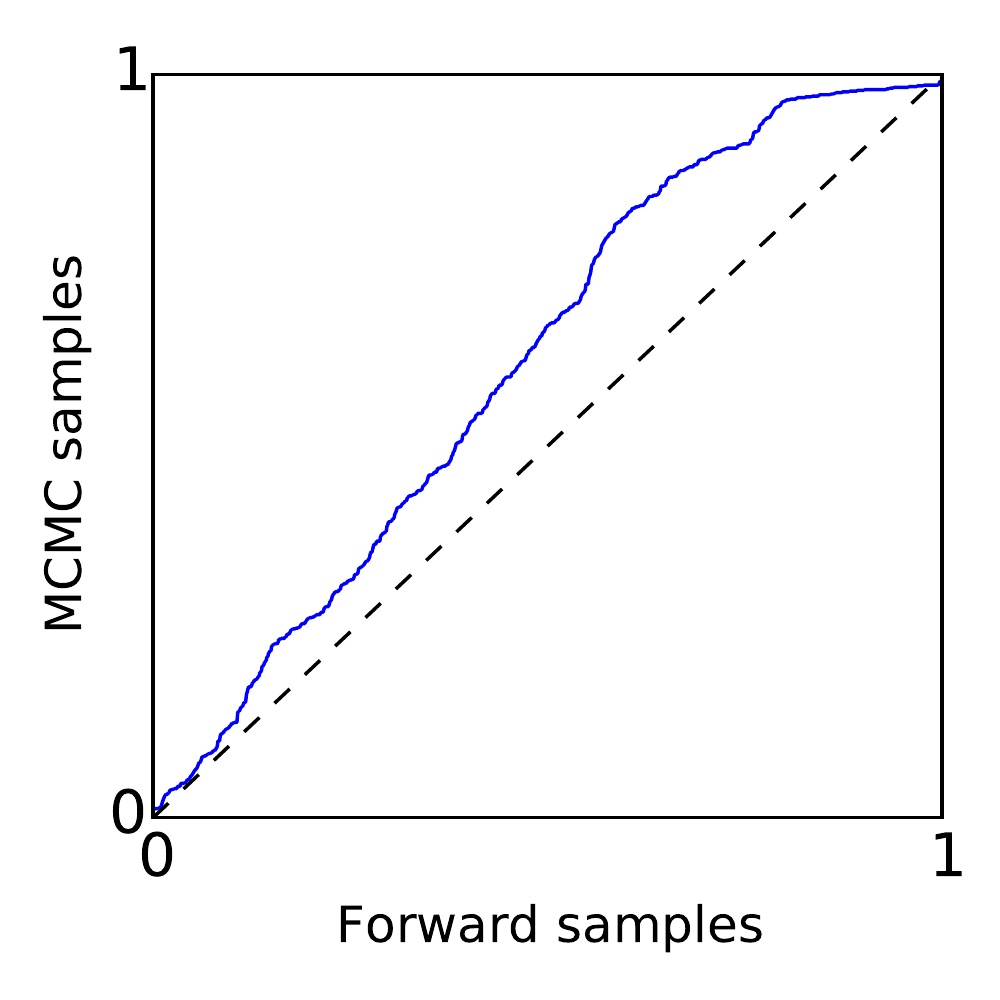}
\end{tabular}
\end{center}
While these figures are synthetic, they're representative of outputs we have seen in the past. In the leftmost plot, the two distributions are indistinguishable, so the test passes. In the middle plot, the distributions are clearly different, so the test fails. In the rightmost plot, the results are noisy and unclear, so the test should be re-run with more forward samples and/or more iterations. In this third case, it may also help to reduce the number of data points. Not only does this speed up each iteration, but it also helps the chains mix faster---the fewer data points there are, the weaker the coupling between $\theta$ and $x$.

Here is a function implementing the Geweke test for the mixture of Gaussians example, where the statistic being compared is $\sigma^2_n$. (In practice, we would compare many more statistics.) Note that this test also requires writing two additional functions to forward sample from the model and to sample from $p(x | \theta)$.

\begin{scriptsize}
\begin{Verbatim}[commandchars=\\\{\}]
\PY{k}{def} \PY{n+nf}{geweke}\PY{p}{(}\PY{n}{num\PYZus{}samples}\PY{p}{)}\PY{p}{:}
    \PY{n}{model} \PY{o}{=} \PY{n}{random\PYZus{}model}\PY{p}{(}\PY{p}{)}

    \PY{n}{forward\PYZus{}results} \PY{o}{=} \PY{p}{[}\PY{p}{]}
    \PY{k}{for} \PY{n}{i} \PY{o+ow}{in} \PY{n+nb}{range}\PY{p}{(}\PY{n}{num\PYZus{}samples}\PY{p}{)}\PY{p}{:}
        \PY{n}{state}\PY{p}{,} \PY{n}{X} \PY{o}{=} \PY{n}{model}\PY{o}{.}\PY{n}{forward\PYZus{}sample}\PY{p}{(}\PY{n}{N}\PY{p}{,} \PY{n}{D}\PY{p}{)}
        \PY{n}{forward\PYZus{}results}\PY{o}{.}\PY{n}{append}\PY{p}{(}\PY{n}{state}\PY{o}{.}\PY{n}{sigma\PYZus{}sq\PYZus{}n}\PY{p}{)}

    \PY{n}{gibbs\PYZus{}results} \PY{o}{=} \PY{p}{[}\PY{p}{]}
    \PY{k}{for} \PY{n}{i} \PY{o+ow}{in} \PY{n+nb}{range}\PY{p}{(}\PY{n}{num\PYZus{}samples}\PY{p}{)}\PY{p}{:}
        \PY{n}{model}\PY{o}{.}\PY{n}{gibbs\PYZus{}step}\PY{p}{(}\PY{n}{state}\PY{p}{,} \PY{n}{X}\PY{p}{)}
        \PY{n}{X} \PY{o}{=} \PY{n}{model}\PY{o}{.}\PY{n}{cond\PYZus{}X}\PY{p}{(}\PY{n}{state}\PY{p}{)}\PY{o}{.}\PY{n}{sample}\PY{p}{(}\PY{p}{)}
        \PY{n}{gibbs\PYZus{}results}\PY{o}{.}\PY{n}{append}\PY{p}{(}\PY{n}{state}\PY{o}{.}\PY{n}{sigma\PYZus{}sq\PYZus{}n}\PY{p}{)}

    \PY{n}{pp\PYZus{}plot}\PY{p}{(}\PY{n}{forward\PYZus{}results}\PY{p}{,} \PY{n}{gibbs\PYZus{}results}\PY{p}{)}
\end{Verbatim}
\end{scriptsize}

Let's now pretend now that we didn't unit test the code, and therefore never caught the typo in \verb+cond_sigma_sq_n+ (where 0.5 was replaced with 0.51). In a practical setting, the effect of this bug would be to make the within-cluster variance about 2\% too large. This is a subtle enough effect that we might never notice it if we simply measured the model's performance on a task, or looked at its predictions.
The beauty of the Geweke test is that it amplifies subtle bugs such as this one. After the parameters are resampled, the data are generated with a 2\% larger noise variance. When the parameters are resampled yet again, the noise variance becomes 2\% larger on top of that, or 4\% in total. Pretty soon, the within-cluster variance explodes:
\begin{center}
\includegraphics[width=0.5\textwidth]{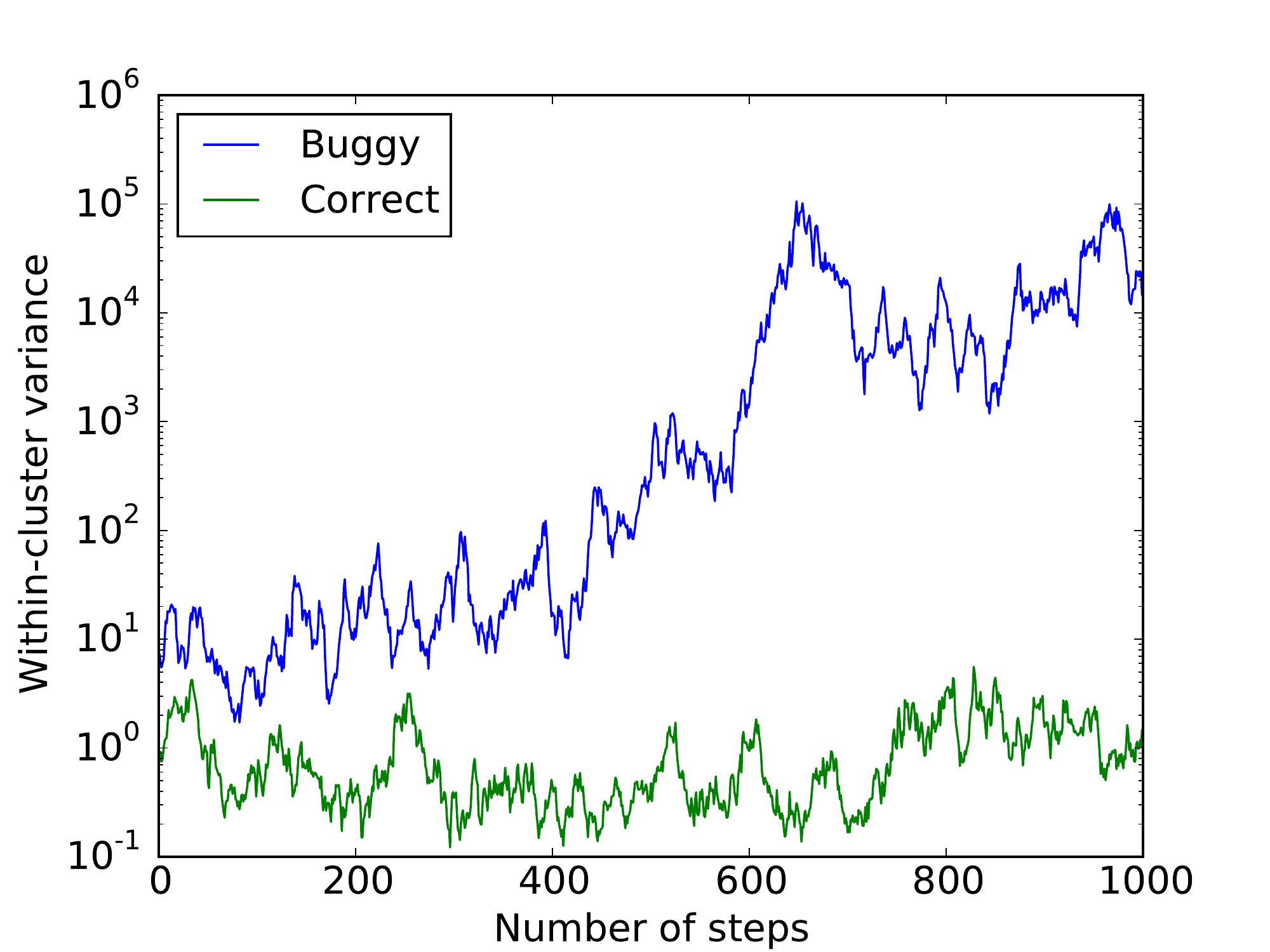}
\end{center}

There's a major drawback of the Geweke test, unfortunately: it gives no indication of where the bug might be. All we can do is keep staring at the code until we find the bug. Therefore, it is best to first ensure all the unit tests pass before running the Geweke test. This meshes with the general test-driven development methodology, where one gets the unit tests to pass before the integration tests.

\section{Discussion}

We have outlined some simple methods for testing MCMC inference code. These methods require writing the code in a modular way, so that different components can be tested independently. If this modular organization is adopted, then the additional work required for unit and integration testing is minimal: it requires writing only two short functions to forward sample from the model and evaluate joint log probabilities, as well as a handful of short testing routines.

As a running example, we discussed a mixture of Gaussians model with isotropic covariance (essentially a Bayesian analogue of k-means). While this model is simple, the testing methodology applies to more complex models and inference algorithms. For instance, some ways one might wish to extend the model and sampler include:
\begin{enumerate}
\item Replace the isotropic covariance with a general covariance matrix (with an inverse Wishart prior)
\item Untie the covariance between different clusters
\item Collapse out the mixture probabilities and the cluster centers in order to speed up mixing
\item Replace the finite Dirichlet mixture with a Chinese restaurant process \citep[\emph{e.g.}][]{infinite-gmm}
\item Use split-merge proposals \cite{split-merge} to speed up mixing
\end{enumerate}
The first two modifications introduce no new difficulties as far as testing. Collapsing out the centers requires defining the collapsed joint probability which would be plugged into (\ref{eqn:consistency}). Furthermore, efficient implementation would require keeping track of sufficient statistics. One can define a data structure which efficiently updates the statistics, and this data structure can be tested entirely using standard testing methodology. Split-merge proposals are defined in terms of conditional probability computations, which can be unit tested the same way as our Gibbs computations; then, the Geweke test can be applied without modification.
A thorough testing framework may even make it easier to add these additional features incrementally, as one can be more confident in the correctness of each modification.

\bibliographystyle{plain}
\bibliography{testing_biblio}

\end{document}